\pdfoutput=1
\documentclass[aps,prd,tightenlines,floatfix,twocolumn,superscriptaddress,showpacs]{revtex4}
\usepackage{graphicx}
\usepackage{wrapfig}
\usepackage{hyperref}
\usepackage[all]{xy}
\usepackage[usenames]{color}

\newcommand{\nc}{\newcommand}
\nc{\beq}{\begin{equation}}
\nc{\eeq}{\end{equation}}
\nc{\bea}{\begin{eqnarray}}
\nc{\eea}{\end{eqnarray}}
\nc{\n}{\nonumber \\}

\nc{\physrep}{Physics Reports}

\begin{document}  

\title{Systematic errors in the measurement of neutrino masses due to baryonic feedback processes: Prospects for stage IV lensing surveys}

\author{Aravind Natarajan}
\email{anat01@me.com}
\affiliation{Department of Physics and Astronomy \& Pittsburgh Particle physics, Astrophysics and Cosmology Center, University of Pittsburgh, 100 Allen Hall, 3941 O'Hara Street, Pittsburgh, PA 15260, U.S.A.} 

\author{Andrew R. Zentner}
\affiliation{Department of Physics and Astronomy \& Pittsburgh Particle physics, Astrophysics and Cosmology Center, University of Pittsburgh, 100 Allen Hall, 3941 O'Hara Street, Pittsburgh, PA 15260, U.S.A.} 
 
\author{Nicholas Battaglia}
\affiliation{McWilliams Center for Cosmology, Department of Physics, Carnegie Mellon University, 5000 Forbes Avenue, Pittsburgh, PA 15213, U.S.A.} 

\author{Hy Trac}
\affiliation{McWilliams Center for Cosmology, Department of Physics, Carnegie Mellon University, 5000 Forbes Avenue, Pittsburgh, PA 15213, U.S.A.}

\begin{abstract}
\noindent We examine the importance of baryonic feedback effects on the matter power spectrum on small scales, and the implications for the precise measurement of neutrino masses through gravitational weak lensing. Planned large galaxy surveys such as the Large Synoptic Sky Telescope (LSST) and Euclid are expected to measure the sum of neutrino masses to extremely high precision, sufficient to detect non-zero neutrino masses even in the minimal mass normal hierarchy. We show that weak  lensing of galaxies while being a very good probe of neutrino masses, is extremely sensitive to baryonic feedback processes. We use publicly available results from the Overwhelmingly Large Simulations (OWLS) project to investigate the effects of active galactic nuclei feedback, the nature of the stellar initial mass function, and gas cooling rates, on the measured weak lensing shear power spectrum. Using the Fisher matrix formalism and priors from CMB+BAO data, we show that when one does not account for feedback, the measured neutrino mass may be substantially larger or smaller than the true mass, depending on the dominant feedback mechanism, with the mass error $|\Delta m_\nu|$ often exceeding the mass $m_\nu$ itself. We also consider gravitational lensing of the cosmic microwave background (CMB) and show that it is not sensitive to baryonic feedback on scales $\ell < 2000$, although CMB experiments that aim for sensitivities $\sigma(m_\nu) < 0.02$ eV will need to include baryonic effects in modeling the CMB lensing potential. A combination of CMB lensing and galaxy lensing can help break the degeneracy between neutrino masses and baryonic feedback processes. We conclude that future large galaxy lensing surveys such as LSST and Euclid can only measure neutrino masses accurately if the matter power spectrum can be measured to similar accuracy.

\end{abstract}
\pacs{98.80.-k, 95.30.Sf, 98.62.Sb, 95.85.Ry}

\maketitle

\section{Introduction}

The discovery of neutrino masses provides exciting hints of physics beyond the standard model. The Sudbury Neutrino Observatory (SNO) \cite{snolab1, snolab2} has detected solar neutrinos at high significance through charged current, neutral current, and elastic scattering reactions, providing strong evidence for neutrino oscillations, and hence for non-zero neutrino masses. This solar neutrino oscillation explained by the Mikheyev-Smirnov-Wolfenstein (MSW) effect \cite{W,MS} is also manifested in the form of a deficit of reactor anti-neutrinos measured by the KamLAND collaboration \cite{kamland1, kamland2}.  On the other hand, analysis of atmospheric neutrinos by Super-Kamiokande \cite{superk1, superk2} also shows evidence for neutrino oscillations, but implies a much larger squared mass difference. The combined data yields a squared mass difference   $\Delta m^2_{21} = 7.58^{+0.22}_{-0.26} \times 10^{-5}$ ev$^2$, and $\Delta m^2_{31} = 2.35^{+0.12}_{-0.09} \times 10^{-3}$ eV$^2$ \cite{prev}. Together, these measurements imply the existence of at least two massive neutrinos with two possible mass orderings. In the normal hierarchy, two neutrinos are nearly degenerate in mass and much lighter than the third neutrino, with a total mass $\Sigma m_\nu \gtrsim 0.058$ eV. The inverted hierarchy has two neutrinos nearly degenerate in mass, and much heavier than the third neutrino. The sum of neutrino masses in this case is $\Sigma m_\nu \gtrsim 0.089$ eV.

Cosmology provides an upper bound to the sum of neutrino masses through precise measurements of the power spectrum of matter fluctuations and the cosmic microwave background (CMB) anisotropies. Neutrinos decoupled from thermal equilibrium while still relativistic, and  constitute hot dark matter. Large neutrino masses therefore result in a damping of the small-scale matter power spectrum. They also modify the size of the sound horizon at decoupling, and can be constrained through the location of the CMB peaks, and Baryonic Acoustic Oscillation (BAO) measurements. Cosmology now provides stringent constraints on neutrino masses. Data from the Atacama Cosmology Telescope (ACT) in combination with WMAP-7, BAO, and Hubble parameter measurements yields a bound on the sum of neutrino masses $\sum m_\nu < 0.39$ eV \cite{act_mnu}. Galaxy angular power spectrum data from the Canada-France-Hawaii-Telescope Legacy Survey was used to place bounds $\sum m_\nu < 0.29$ eV at 95\% confidence \cite{cfhtlens}. The Planck collaboration obtained a limit on the sum of masses $\Sigma m_\nu < 0.23$ eV at the 95\% confidence level \cite{planck} using CMB+BAO data. When Lyman-$\alpha$ data is included, Ref. \cite{seljak} found an upper limit $\Sigma m_\nu < 0.17$ eV at 95\% confidence. Authors \cite{sorensen} also found an upper limit of $\Sigma m_\nu < 0.17$ eV at 95\% confidence using Planck+BAO+HST+WiggleZ data. 

In addition to the aforementioned constraints, two new analyses have yielded tentative indications of neutrino masses based on cosmological signatures. The South Pole Telescope (SPT) \cite{spt_neu} reported the detection of non-zero neutrino masses at the $3 \sigma$ level using CMB+BAO+$H_0$+SPT$_{\rm CL}$ data, favoring a mass sum $\sum m_\nu = 0.32 \pm 0.11$ eV. More recently, the Sloan Digital Sky Survey \cite{sloan_nu} found results favoring a neutrino mass sum $\sum m_\nu = 0.36 \pm 0.10$ eV from the Baryon Oscillation Spectroscopic Survey (BOSS) CMASS Data Release 11, in good agreement with the results from the SPT experiment. If these exciting results are confirmed by future experiments, they would have major implications for both particle physics and cosmology. It is therefore important to take a critical look at the difficulties facing neutrino mass measurements from cosmological surveys.

Upcoming surveys such as the Large Synoptic Sky Telescope (LSST) and the Euclid mission are expected to substantially improve our understanding of neutrino physics. Combining weak lensing shear constraints from LSST with Planck constraints, one could expect to measure neutrino masses down to $m_\nu \sim$ 0.03 eV and $\Delta N_{\rm eff} \sim 0.08$ \cite{lsst_book}, where $N_{\rm eff}$ is the number of effective neutrino like degrees of freedom. Future surveys such as the Euclid mission may obtain even stronger constraints on the neutrino mass. Ref.~\cite{very_optimistic_people} estimate that combined CMB, shear, and galaxy data from future surveys can constrain neutrino masses with an estimated error on the sum of neutrino masses $\sigma(m_\nu) < 0.011 (0.022)$ eV, assuming full knowledge (no knowledge) of the galaxy bias, which would be a $\gtrsim 2.6 \sigma$ detection even in the case of the minimal mass normal hierarchy.  

Such highly precise measurements require a thorough understanding of the clustering properties of matter through high-resolution hydrodynamic simulations. Dark matter only simulations can attain sub-percent accuracy only on quasi-linear scales $k \sim 1$ $h$/Mpc (see, for example \cite{coyote}, but also the recent extension \cite{coyote_ext}). However, these predictions cannot be used without modification to undertake precision cosmology with weak lensing. On scales $k \gtrsim 0.1\, h^{-1}\mathrm{Mpc}$, baryonic effects can modify the matter power spectrum at levels that are large compared to the expected precision with which weak lensing will be measured \cite{jing_etal, rudd_etal, semb1, semb2, vand1} and can also alter the observed clustering of galaxies \cite{vand2}. In this article, we investigate how the sensitivity of neutrino mass measurements is affected by baryonic processes. In Section II, we discuss baryonic feedback processes using results from the  Overwhelmingly Large Simulations (OWLS) project \cite{owls}. In Section III, we compute the shear power spectrum and show that ignoring feedback effects can substantially bias the estimated neutrino mass. We also consider CMB lensing and show that unlike weak lensing of large scale structure, it is practically unaffected by baryonic feedback processes. Finally, we present our conclusions.

\section{Neutrinos, baryonic effects, and the matter power spectrum on small scales}

Neutrinos make up  a  small percentage of dark matter, but they are  hot, meaning that they are relativistic at freeze-out. The neutrino number density $n_\nu$ at the present epoch is:
\beq
n_\nu = \frac{3}{4} n_{\rm cmb} \left( \frac{T_\nu}{T_{\rm cmb}} \right )^3 = \frac{3}{11}n_{\rm cmb} \sim 112 \; {\rm cm}^{-3},
\eeq
where $n_{\rm cmb}$ is the present day number density of CMB photons with temperature $T_{\rm cmb}$, and $T_\nu = (4/11)^{1/3} \; T_{\rm cmb}$ is the neutrino temperature.   The present day energy density of neutrinos, assuming $m_\nu \gg T_{\rm cmb}$ is then
\bea
\rho_\nu^{\rm norm} &=&  \frac{2N_{\rm eff}}{3} \frac{7}{8} \left( \frac{4}{11} \right )^{4/3}\rho_{\rm cmb} + \frac{N_{\rm eff}}{3} n_\nu m_\nu \n
&\approx&   \left [ 1.14 \times 10^{-5} + 1.08 \times 10^{-3} \left( \frac{m_\nu}{0.1 \, {\rm eV}} \right ) \right ] \; {\rm eV}/{\rm cm}^3 \;\;\;\;\;\;
\label{norm}
\eea
for the minimal normal hierarchy with one massive neutrino and two (nearly) degenerate massless neutrinos. If instead, neutrino masses followed the minimal inverted hierarchy scenario, we would find
\beq
\rho_\nu^{\rm inv} = \frac{N_{\rm eff}}{3} \frac{7}{8} \left( \frac{4}{11} \right )^{4/3}\rho_{\rm cmb} + \frac{2N_{\rm eff}}{3} n_\nu m_\nu \n,
\label{inv}
\eeq
where we have assumed two degenerate neutrino masses, and one massless neutrino. Thus whenever the first term in Eq. [\ref{norm}] or Eq. [\ref{inv}] may be neglected, $\rho_\nu$ remains the same provided $m_\nu \rightarrow \sum m_\nu$. As expected, current bounds on neutrino masses are not very sensitive to the assumed hierarchy \cite{sorensen}. Nevertheless, it is intriguing that highly precise future observations may be able to determine the neutrino hierarchy \cite{verde}. $N_{\rm eff}$ is the number of relativistic, neutrino-like  degrees of freedom. Constraints on $N_{\rm eff}$ from CMB+BAO \cite{planck} give us $N_{\rm eff} = 3.30^{+0.54}_{-0.51}$. Precision electroweak measurements of the decay of the Z boson yield the number of light neutrino-like species = 2.9840 $\pm$ 0.0082, in agreement with the existence of 3 species of neutrinos \cite{ew}.  Here, we assume $N_{\rm eff} = 3.046$ as expected in the standard model (Note that $N_{\rm eff}$ is slightly larger than 3 due to heating of neutrinos at the time of $e^+e^-$ annihilation).

\begin{figure*}[!t]
\begin{center}
\scalebox{1.5}{\includegraphics{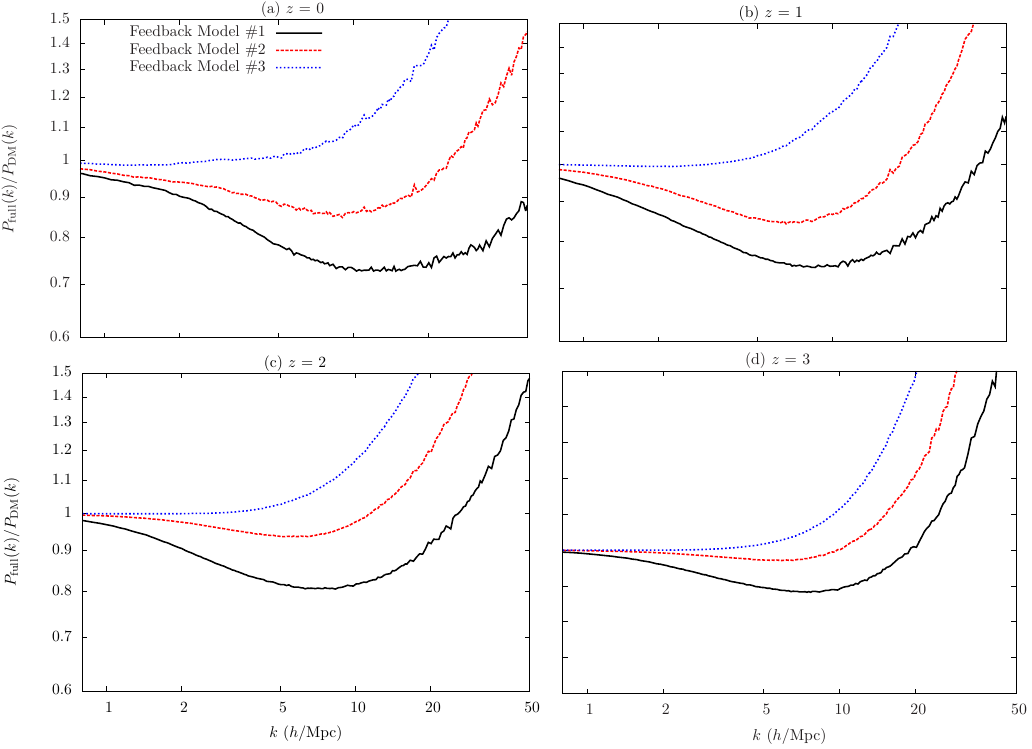}}
\end{center}
\caption{ The square of the baryon bias, $b^2(k) = P_{\rm full}(k)/P_{\rm DM}(k)$, for redshifts $z$ = 0,1,2,3. The black (solid), red (dashed), and blue (dotted) curves show the effect of baryonic feedback on the matter power spectrum, for feedback models \#1, \#2, and \#3 respectively.  Feedback model \#1 includes AGN feedback (labeled {\scriptsize AGN} in \cite{owls}). Feedback model \#2 accounts for a top-heavy IMF and extra SN energy (labeled {\scriptsize DBLIMFV1618} in \cite{owls}), while Feedback model \#3 has no SN feedback and cooling by primordial elements only (labeled {\scriptsize NOSN-NOZCOOL} in \cite{owls}). As expected, the model with AGN feedback (Model \#1) shows the largest damping in the power spectrum on small scales at $z$=0. The model with no metal cooling or SN feedback (Model \#3) shows a boost in the power spectrum on small scales due to contraction of halos. 
\label{fig1} }
\end{figure*}

\emph{Damping the matter power spectrum:}
The effect of including a nonzero mass $m_\nu$ is to cause a damping in the matter power spectrum $P(k)$ on small scales due to free streaming of neutrinos. The deficit in power $\Delta P(k)$ on scales well below the free streaming length, $\lambda_{\mathrm{FS}} \approx  8.6\, h^{-1}\mathrm{Mpc}\, \sqrt{(1+z)(0.238/\Omega_{\rm m})}(\mathrm{eV}/m_{\nu})$, approaches (see for e.g. Ref~\cite{cooray}):
\beq
\frac{\Delta P}{P} \approx -8 \frac{\rho_\nu}{\rho_{\rm m}}.
\eeq
It is also important to account for the effect of neutrinos when one is modeling the non-linear power spectrum (see for e.g. \cite{nlnu1, nlnu2, nlnu3, nlnu4, nlnu5, nlnu6, nlnu7,neu_nl}).
Since weak gravitational lensing is sensitive to the small-scale matter power spectrum, it is an excellent probe of neutrino masses \cite{cooray}. However, we will see that other small-scale effects such as baryonic feedback can mimic the effect of neutrino masses, resulting in a significant error in the measurement of neutrino masses.

It is now well known (see for e.g. \cite{jing_etal, rudd_etal}) that baryonic processes can alter the matter power spectrum on small scales. Authors \cite{jing_etal} numerically computed the matter power spectrum on non-linear scales, and found that clustering of gas is suppressed, and the clustering of dark matter is enhanced on scales $k > 1$ $h$/Mpc. They found that the total matter power spectrum is suppressed by $\sim$ 1\% for $1 \lesssim k \lesssim 10$ $h$/Mpc due to shock heating of the gas and thermal pressure. The total power spectrum was boosted on smaller scales due to halo contraction in response to gas condensation. Ref.~\cite{rudd_etal} reported qualitatively similar results, but with a significantly larger boost in power on scales $k \gtrsim 1\, h \mathrm{Mpc}^{-1}$ and a $\sim 5-10\%$ deficit in power on scales as large as $k \sim 0.1\, h\mathrm{Mpc}^{-1}$. In this article, we consider simulation results from the OWLS project \cite{owls}, which span a wide range of implementations of baryonic processes, to model the effects of baryons on the matter power spectrum. \\

\emph{Simulations and feedback models:}
The OWLS project consists of a large suite of cosmological smoothed particle hydrodynamic (SPH) simulations with varying box sizes and resolutions, using $2 \times 512^3$ particles \cite{owls}. Each simulation is repeated many times with different subgrid prescriptions. The simulations were performed with an extended version of {\scriptsize GADGET-3} \cite{Gadget}, which is a Lagrangian code used to calculate the gravitational evolution of the matter and the hydrodynamics of the baryonic gas. The cosmology used in the OWLS suite is that of WMAP-3: $\{ \Omega_{\rm m} = 0.238, \Omega_{\rm b} = 0.0418, \Omega_\Lambda = 0.762, \sigma_8 = 0.74, n_{\rm s} = 0.951, h = 0.73 \}$. Radiative cooling and heating are implemented. We consider 3 feedback models varying active galactic nuclei (AGN) feedback, star formation, and cooling rate, which provide effects representative of the wide range of effects considered by \cite{owls}. Matter accreting onto supermassive black holes emits enormous amounts of high energy radiation which if coupled to the gas, can significantly alter the clustering of matter on small scales. The AGN feedback model used in the OWLS suite \cite{owls} is described in \cite{booth_schaye}, and involves placing a seed black hole in every dark matter halo whose mass exceeds a certain minimum value $m_{\rm min} = 4 \times 10^{10} M_\odot$. Fifteen percent of the energy radiated by the infalling matter is assumed to couple to the surrounding gas, in order to match the observed cosmic mass density in black holes, as well as the relation between black hole and galaxy mass, both at redshift zero.

Similarly details of star formation and cooling can significantly influence the clustering of matter on small scales. A top-heavy initial mass function (IMF) may be expected in extreme environments such as the galactic center, or in starburst galaxies (see Ref. \cite{owls} and references therein).  When the supernova (SN) energy scales with emissivity of ionizing radiation, Ref. \cite{owls} find that a top-heavy IMF yields 7.3 times more SN energy per unit stellar mass, compared to the Chabrier IMF \cite{chabrier}. The top-heavy IMF also yields more metal mass per stellar mass, which increases metal line cooling rates, which in turn increases the star formation rate. It is also interesting to consider simulations that do not include metal line cooling or SN driven winds. Except at very high redshifts, the absence of metal line cooling suppresses star formation. Authors \cite{owls} find that ignoring metal cooling may decrease the total number of stars by a factor of 2. 

Let us define the \emph{baryon bias} as the ratio of the full matter power spectrum in a simulation that includes baryonic processes to the matter power spectrum predicted by a simulation that considers dark matter only,  
\beq
b^2(z,k) = \frac{P_{\rm full}(z,k)}{P_{\rm DM}(z,k)},
\label{bias}
\eeq
where $P_{\rm full}(k)$ is the matter power spectrum including feedback effects. Fig. \ref{fig1} shows the square of the baryon bias, for three different feedback models, at redshifts $z$ = 0,1,2,3.  Feedback model \#1 includes AGN feedback (labeled {\scriptsize AGN} in \cite{owls}). Feedback model \#2 accounts for a top-heavy IMF and extra SN energy (labeled {\scriptsize DBLIMFV1618} in \cite{owls}), while Feedback model \#3 has no SN feedback and cooling by primordial elements only (labeled {\scriptsize NOSN-NOZCOOL} in \cite{owls}). As expected, the model with AGN feedback (Model \#1) shows the largest damping in the power spectrum on small scales at $z$=0. The model with no metal cooling or SN feedback (Model \#3) shows a boost in the power spectrum on small scales due to contraction of halos. The results seen in the OWLS study show much larger damping than what was estimated by \cite{jing_etal}, particularly for the case of AGN feedback.

\begin{figure*}[t]
\begin{center}
\scalebox{0.6}{\includegraphics{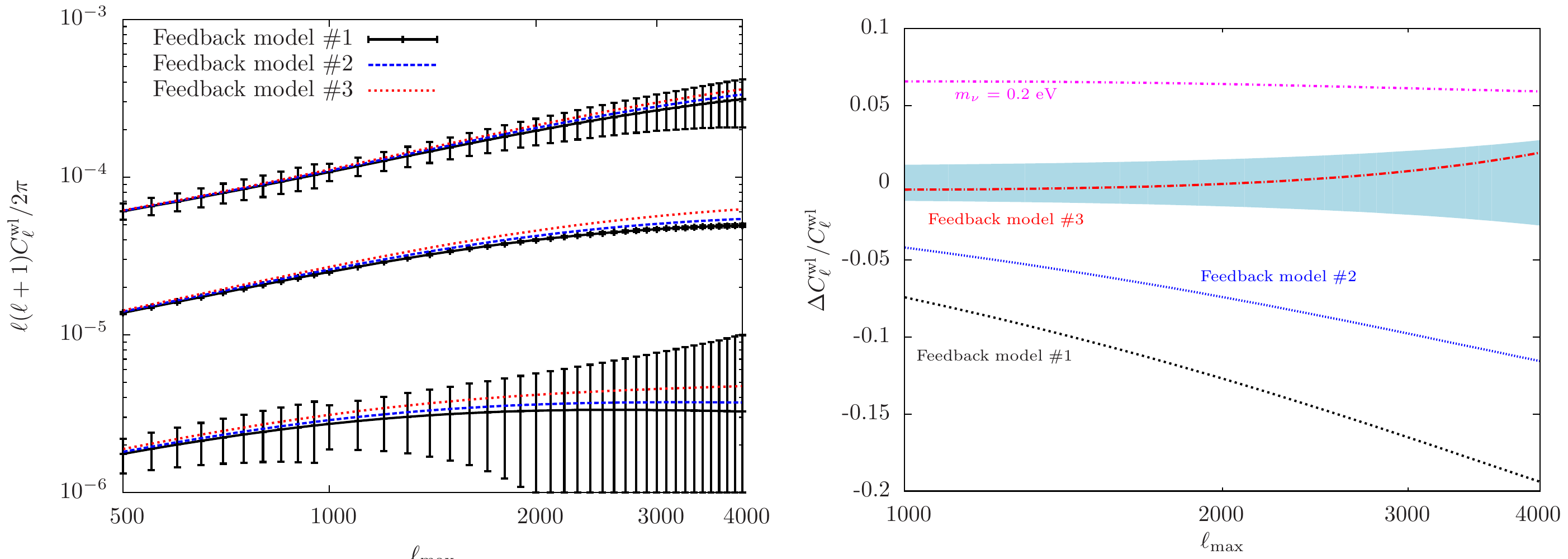}}
\end{center}
\caption{ \emph{Left:} The weak lensing  power spectrum for source redshifts $z$ = 0.2, 1.0, and 3.0, assuming $m_\nu$ = 0.3 eV. Only the auto-power spectra are shown. The black curves assume feedback model \#1, the blue curves are plotted for feedback model \#2, and the red curves for feedback model \#3.  Also shown are expected error bars for an LSST-like or Euclid-like instrument (for feedback model \#1), with bin size $\Delta\ell$ = 50 for $\ell < 1000$, and $\Delta\ell$ = 100 otherwise. The error bars are large for  small $z$ because of low volume, and for  large $z$ because of low flux. \emph{Right:} Fractional change in the shear auto-power spectrum for $z=1.0$: $\Delta C_\ell = [C_\ell - C_\ell ({\rm ref})] / C_\ell({\rm ref})$, where the reference model has no feedback and assumes a neutrino mass of 0.3 eV. The magenta curve considers a smaller neutrino mass $m_\nu$ = 0.2 eV. 
The shaded area represent the error for $z$ = 1.0.
\label{fig2} }
\end{figure*}

The effects of baryonic feedback have also been studied by more recent simulations, e.g. ILLUSTRIS \cite{illustris1, illustris2, illustris3} which considers radiative cooling with self shielding corrections, star formation feedback to drive kinetic galactic winds, black hole seeding, and 3 kinds of AGN feedback: quasar mode, radio mode, and a radiative mode. ILLUSTRIS includes 15 or so free parameters, associated with the various feedback processes. Authors \cite{illustris1, illustris2, illustris3} compare the mean relation between stellar mass and halo mass (with a smaller simulation)  with and without radio mode AGN feedback. There is little difference between the simulations at low halo masses and at high redshifts. However at $z=0$, and halo mass $> 10^{13} M_\odot$, the stellar mass with AGN feedback is less than half the value found in the simulation without feedback,  showing that AGN activity provides strong negative feedback. AGN feedback also influences the dependance of the specific star formation rate with halo mass. It is however noted that at low redshifts, massive halos of $\sim 10^{13} M_\odot$ are almost devoid of gas as a result of radio-mode AGN feedback in disagreement with observations, suggesting that the precise nature of AGN feedback is a difficult issue to resolve.

\section{The weak lensing shear power spectrum}

The bending of light around massive objects results in distortion and magnification of images, quantified by the shear and convergence fields (for a detailed review, see \cite{bartelmann}).  These effects may be used to probe the properties of the matter distribution between us and the source. When the effects are statistical in nature, i.e. when they are apparent only when a large number of sources are present, we are in the regime of weak gravitational lensing. While galaxies are elliptical, they are randomly oriented. Thus, when one averages over a large number of galaxies, the residual ellipticity is a measure of the weak lensing shear. The shear power spectrum considering sources in redshift bins $i$ and $j$ is given by (see for e.g. \cite{bartelmann, huterer}):
\bea
C_{ij}(\ell) &=& \frac{9}{4} \left( \frac{H_0} {c} \right )^3 \Omega^2_{\rm m} \int_0^\infty  \; \frac{dz (1+z)^2}{\sqrt{\Omega_{\rm m}(1+z)^3 + \Omega_\Lambda}} \n
&\times& W_i(z) W_j(z) P\left(0,\frac{\ell}{\chi}\right ) D^2(z) b^2(z,\frac{\ell}{\chi}),
\label{shear}
\eea
where we have set $k = \ell/\chi(z)$, where $\chi(z)$ is the comoving distance:
\beq
\chi(z) = \int_0^z \frac{c \, dz}{H(z)}.
\eeq
$P(0,k)$ is the matter power spectrum at redshift $z$=0, and $H(z)$ is the Hubble parameter. $D(z)$ is the growth function of density perturbations given by
\beq
D(z) = \exp \left[- \int_0^z \frac{dz'}{1+z'}  \Omega_{\rm m}^\gamma(z') \right], 
\eeq
where $\gamma$ = 0.55 \cite{linder}, and the matter density at redshift $z$ is
\beq
\Omega_{\rm m}(z) = \frac{\Omega_{\rm m} (1+z)^3}{\Omega_{\rm m} (1+z)^3 + \Omega_\Lambda}.
\eeq
$b(z,k)$ is the baryon bias as defined in Eq. [\ref{bias}]. The function $W_i(z)$ is given by:
\beq
\label{eq:lensingkernel}
W_i(z) = \int_z^\infty dz' \, n_{\rm gal}(z') \xi(z',z_i) \left[ 1 - \frac{\chi(z)}{\chi(z')} \right ].
\eeq
$n_{\rm gal}(z)$ describes the redshift distribution of the source galaxies normalized so that $\int dz \, n_{\rm gal}(z)$ = 1. We choose the form \cite{chang_etal}:
\beq
n_{\rm gal}(z) = \frac{4}{\sqrt{\pi}} \frac{z^2}{z^3_0} \exp \left[ {-\left( z/z_0 \right )^2} \right ].
\eeq
The function $n_{\rm gal}(z)$ tells us that most galaxies are observed around $z \sim z_0$, i.e. far enough to cover a significant volume, yet close enough to be visible with the telescope. $\xi(z_s,z)$ is a suitably chosen window function for the source redshift bin $z_s$. We pick a top hat window function which is 1.0 within the bin and zero outside.

The observed power spectra $P_{ij}(\ell)$ contain both signal and shot noise components:
\beq
P_{ij}(\ell) = C_{ij}(\ell) + \delta_{ij}\frac{ \sigma^2_\epsilon}{n_i},
\label{eqnP}
\eeq
where $\sigma_\epsilon$ is the intrinsic ellipticity of galaxies, and $n_i$ is the number of galaxies present in the redshift bin $i$. 

We compute the shear power spectrum, assuming the following cosmology: $\{ \Omega_{\rm b}h^2 = 0.0222, \Omega_{\rm c}h^2 = 0.118, h = 0.674, 10^9A_{\rm s} = 2.21, n_{\rm s} = 0.962 \}$, where $\Omega_{\rm b}$ and $\Omega_{\rm c}$ are the present day baryon and cold dark matter density fractions, $h$ is the hubble parameter in units of 100 km/s/Mpc, $A_{\rm s}$ is the amplitude of the primordial scalar curvature power spectrum, and $n_{\rm s}$ is the scalar spectral index.

We use the bias rather than the power spectrum, so it is less sensitive to cosmology. $\sigma_8$ does not change the overall amplitude of the bias, but just the small-scale shape, which depends on the nonlinear collapsed fraction. Our cosmology accounts for the most recent estimate of $\sigma_8$ based on Planck + ACT + SPT + BAO, i.e. $\sigma_8 = 0.826 \pm 0.012$, and is over $5 \sigma$ larger than the value used in the OWLS simulations. The smaller value of $\sigma_8$ used by OWLS results in a smaller number of halos being present at a given redshift, and hence smaller baryonic feedback. We therefore consider the OWLS simulation results to be a conservative estimate for baryonic feedback. We believe that using simulations with a more realistic value of $\sigma_8$ would make our results even more relevant. Also, the signal-to-noise ratio increases with $\sigma_8$ so that increasing $\sigma_8$ leads to slightly stronger constraints.

Fig. \ref{fig2}  shows the auto power spectra for redshifts $z$ = 0.2, 1.0, and 3.0, for the 3 feedback models considered in Fig. \ref{fig1}. Also shown are the error bars expected for LSST, plotted for  feedback model \#1. For the cosmic variance component, we choose $f_{\rm sky} = 0.5$, approximately equal to 20,000 square degrees of sky coverage. For the shot noise term, we choose $n$ = 50 galaxies per square arcminute, with $\sigma_\epsilon$ = 0.22 \cite{lsst_book, lsstwww}. We assume the median redshift of the survey = 1.0, which sets $z_0$ = 0.92 \cite{chang_etal}. It is clear that the observations can easily distinguish between feedback models for $z=1$. Conversely, if the true model is unknown, there will substantial errors in the inferred cosmological parameters. For $z \lesssim 0.2$, the error bars are large since the volume covered is very small. Similarly, for $z \gtrsim 3.0$, the error bars are similarly large since the sample of galaxies is flux limited.

\section{Results}

Let us now estimate the errors in the neutrino mass measurement using the Fisher matrix formalism.
Let  $\vec\theta = \{ \Omega_{\rm b}h^2, \Omega_{\rm c}h^2, h, 10^9A_{\rm s}, n_{\rm s}, m_\nu \}$ be the set of cosmological parameters to be constrained. The Fisher matrix is then
\beq
{\bf F} = {\bf C_{\rm prior}^{-1}} + \sum_\ell \; \frac{\partial {\bf P}}{\partial \vec {\theta}} \; {\bf Cov}^{-1} \; \frac{\partial {\bf P^{\rm T}}}{\partial \vec {\theta}}.
\eeq

\begin{figure}[t]
\begin{flushleft}
\scalebox{0.6}{\includegraphics{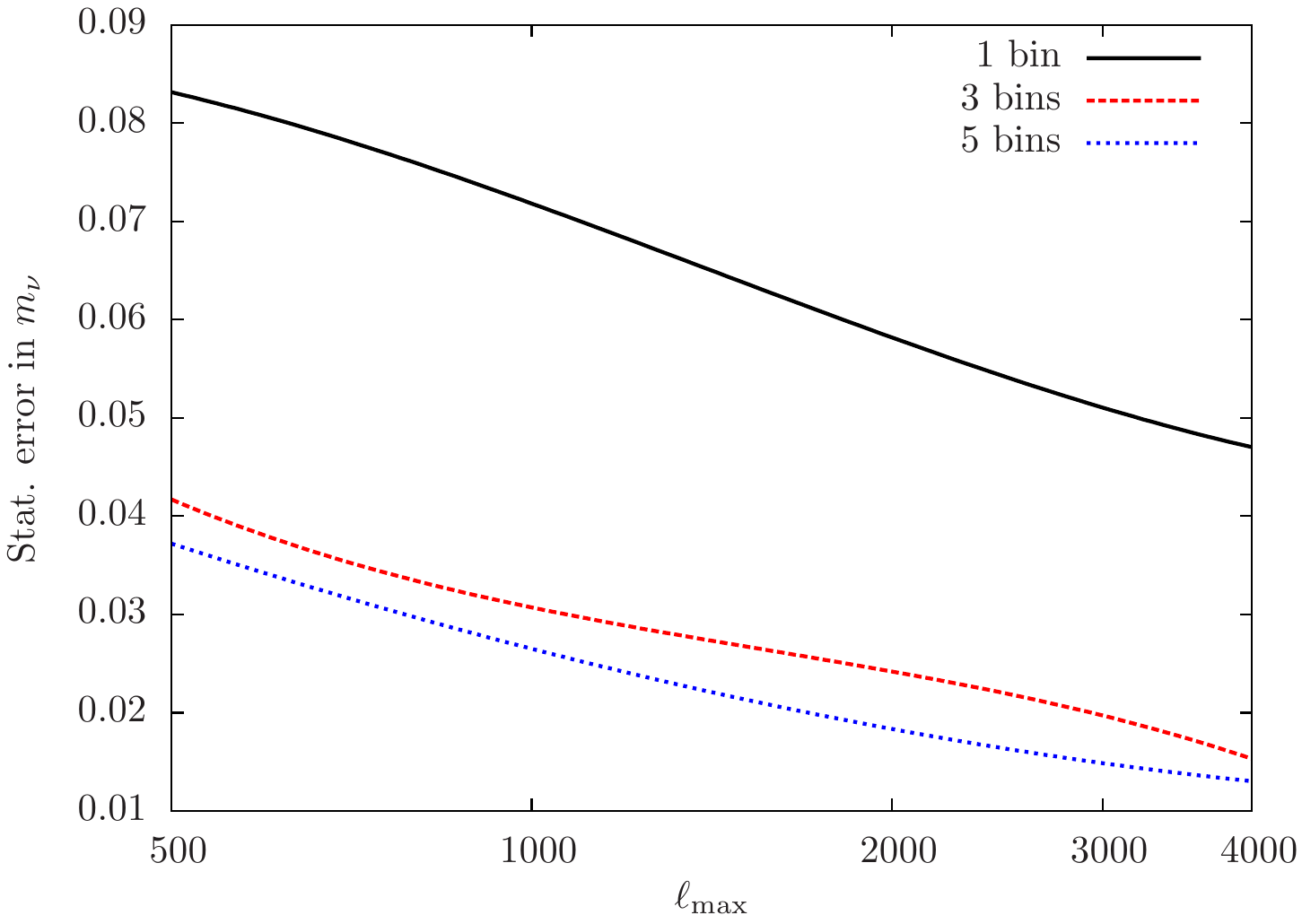}}
\end{flushleft}
\caption{ Expected statistical error in the measurement of neutrino masses, assuming multipoles up to $\ell_{\rm max}$ can be measured. The redshift range is divided into 1 bin (black, solid) of size $\Delta z = 2.5$, 3 bins (red, dashed) with each bin of size $\Delta z = 0.8$, and 5 bins (blue dotted) with bin size $\Delta z = 0.5$. There is one power spectrum with 1 bin, six power spectra with 3 bins, and fifteen power spectra with 5 bins. 
\label{fig3} }
\end{figure}

\begin{figure*}[t]
\begin{center}
\scalebox{0.65}{\includegraphics{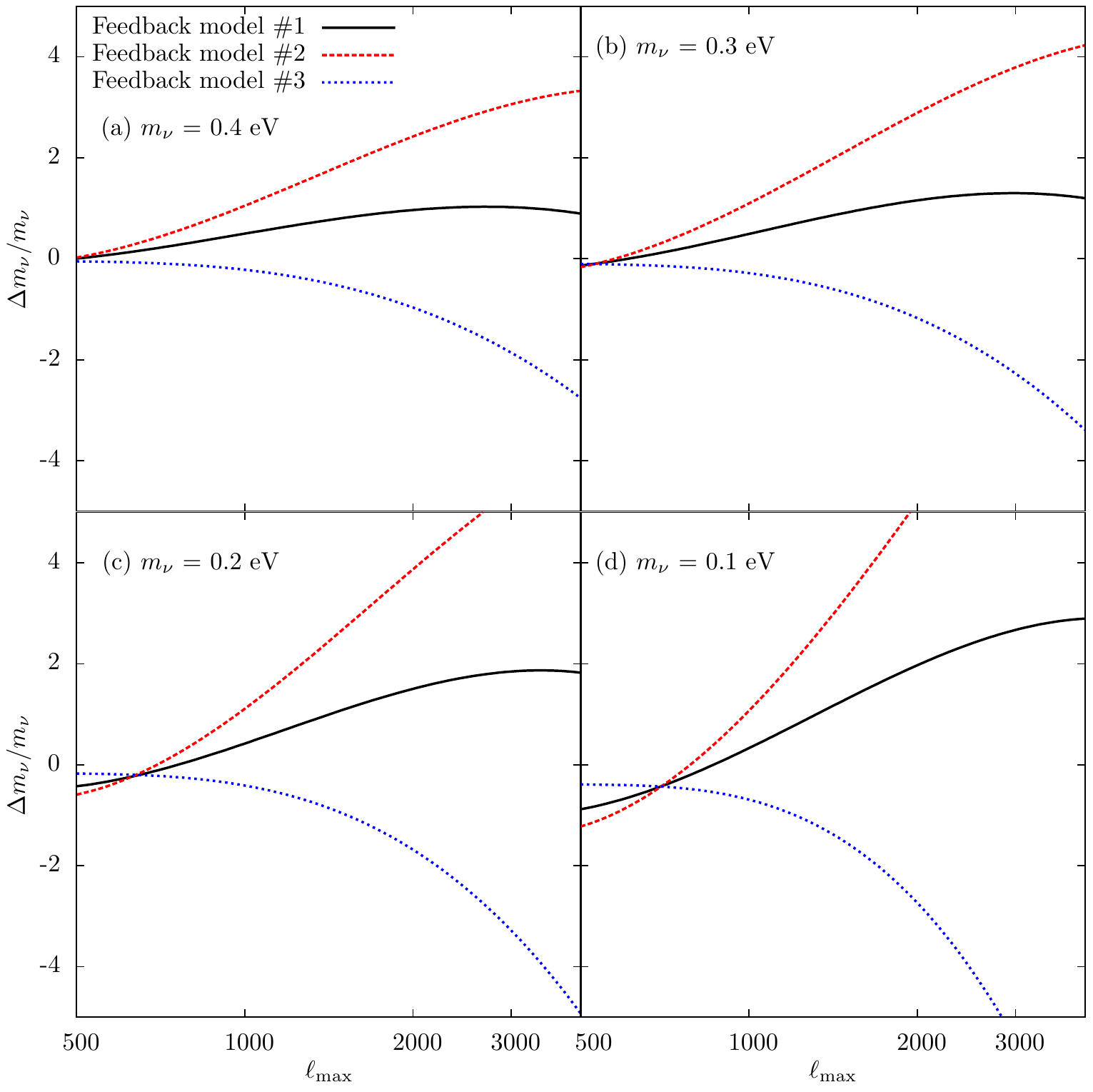}}
\end{center}
\caption{ Bias in the estimated neutrino mass from weak lensing assuming LSST parameters, relative to the assumed neutrino mass.  Source galaxies in the redshift range $0<z<2.5$ are divided into 5 bins. The bias, or systematic error is due to ignoring baryonic feedback. The four panels show the bias $\Delta m_\nu$ for assumed neutrino masses $m_\nu$ = 0.4, 0.3, 0.2, 0.1 eV. The 3 curves are plotted for feedback models \#1, \#2, and \#3.  Note that the Fisher matrix formalism is not reliable when $|\Delta m_\nu| > m_\nu$.
\label{fig4} }
\end{figure*}
${\bf C_{\rm prior}}$ is the covariance matrix obtained from CMB+BAO data, and serves to place priors on all cosmological parameters. ${\bf P}^{\rm T}$ indicates the transpose of ${\bf P}$ (see Eq. [\ref{eqnP}]), and ${\bf Cov}$ is the lensing covariance matrix defined by
 \beq
{\rm Cov}_{AB}(P_{ij},P_{kl}) = \frac{2}{(2\ell+1)f_{\rm sky}\Delta\ell} \, \left[ \frac{P_{ik}P_{jl} + P_{il}P_{jk}}{2} \right ].
\eeq
In order to map the power spectrum indices $(i,j,k,l)$ on to covariance matrix indices $(A,B)$, we use the following rule: $A(i,j) = {1,2,3,4,\cdots}$ for $(i,j) = (1,1), (2,1), (2,2), (3,1) \cdots$, and similarly for $B(k,l)$.  Thus, when the data consists of only 1 redshift bin, the covariance matrix is $1\times1$. With 2 bins, one can construct 2 auto-power spectra, and 1 cross-power spectrum. The covariance matrix is then $3 \times 3$. With 5 bins, one can construct 15 power spectra, and {\bf Cov} is a $15\times15$ matrix. Using the Fisher matrix formalism, one may obtain an estimate of the statistical error on each cosmological parameter $\theta_i$: $\sigma(\theta_i) = \sqrt{  \left[ {\bf F}^{-1} \right ]_{ii} }$, where ${\bf F}^{-1}$ is the inverse of the Fisher matrix.  

Fig. \ref{fig3} shows the estimated statistical error on the neutrino mass $m_\nu$, if multipoles up to $\ell_{\rm max}$ can be measured. We bin multipoles in steps of $\Delta\ell$ = 50 for $\ell < 1000$ and $\Delta\ell$ = 100 for $\ell > 1000$. Shown are results for $n$ = 1, 3, and 5 redshift bins. When $n$ = 1, we do no tomography, and include all galaxies in the range $0<z<2.5$ in the same bin (bin size $\Delta z$ = 2.5). When $n$ = 3, we include galaxies in the range $0<z<2.4$ in 3 bins of width $\Delta z$ = 0.8. When $n$ = 5, we consider all galaxies in the range $0<z<2.5$ divided equally into 5 bins with bin size $\Delta z$ = 0.5. It is clear that the ability to probe redshift evolution through the large number of auto and cross power spectra for $n$ = 3 and $n=5$ substantially improves the sensitivity of the experiment. The constraining power of weak lensing tomography on neutrino mass is not greatly improved for $n>5$
due to the fundamentally poor resolution of redshift-dependent effects induced by the broad lensing kernel in Eq.~(\ref{eq:lensingkernel}). Very few galaxies beyond redshift of $z \sim 2.2$ are likely to be exploited for weak lensing, and including or excluding galaxies with $z \gtrsim 2.2$ has very little effect on cosmological constraints \cite{ma_etal, hearin_etal}. For simplicity, we have neglected photometric redshift errors in this calculation.


The potential for weak lensing tomography to constrain neutrino mass, as shown in Fig.~\ref{fig3} is impressive. In particular, our results suggest that the statistical errors on the sum of the neutrino masses for a Stage IV, large-scale imaging survey is on the order of $\sigma(m_{\nu}) \sim 0.015\, \mathrm{eV}$, in agreement with previous work \cite{snowmass_people}. This level of constraint is clearly sufficient to detect neutrino masses of the level expected based upon oscillation data, and can even isolate the neutrino mass hierarchy.

Let us now consider the bias introduced in our measurement of parameter $\theta_i$ when baryonic feedback is ignored. Let $\Delta C_\ell$ be the difference between the true lensing power spectrum (which accounts for the correct feedback model) and the assumed faulty lensing power spectrum for a clustering model that accounts only for baryons and cold dark matter but ignores baryonic small-scale physics such as AGN, star formation, and cooling. The estimated cosmological parameters will all be offset from the true values by an amount $\Delta\vec\theta$ given by:
\beq
\Delta\vec\theta = {\bf F}^{-1} \sum_\ell \, \frac{ \partial {\bf C}_\ell}{\partial\theta}  \, {\bf Cov}^{-1} \, \Delta {\bf C}_\ell
\label{bias}
\eeq

Fig. \ref{fig4} shows the bias or systematic error in the estimated neutrino mass relative to the assumed value of $m_\nu$ when baryonic effects are ignored. The four panels show the bias as a fraction of the true neutrino mass for $m_\nu$ = 0.4, 0.3, 0.2, and 0.1 eV. The black (solid) curve is plotted when the true feedback model is that of \#1. The red (dashed) and blue (dotted) are drawn for feedback models \#2 and \#3 respectively. The bias becomes more significant as $\ell_{\rm max}$ is increased, since small scales are more affected by baryonic physics. In view of this, it may be preferable to restrict ourselves to low $\ell_{\rm max}$, at the expense of larger statistical errors. 

The bias in the neutrino mass can be either positive or negative depending on the scale at which the baryon bias reaches its lowest value. Thus with feedback models \#1 and \#2, we overestimate the neutrino mass, but with feedback model \#3, we underestimate it. The bias is substantial: with error bars representative of LSST or Euclid, we obtain bias values $|\Delta m_\nu| \gg m_\nu$. Unfortunately, the Fisher matrix formalism is not applicable in this regime since the Fisher matrix is obtained through Taylor series expansion of the logarithm of the likelihood function about its maximum value. We can nevertheless infer that the bias is significant and a careful understanding of baryonic physics is essential if cosmological experiments are to obtain the correct neutrino mass.

\begin{figure}[!b]
\begin{center}
\scalebox{0.7}{\includegraphics{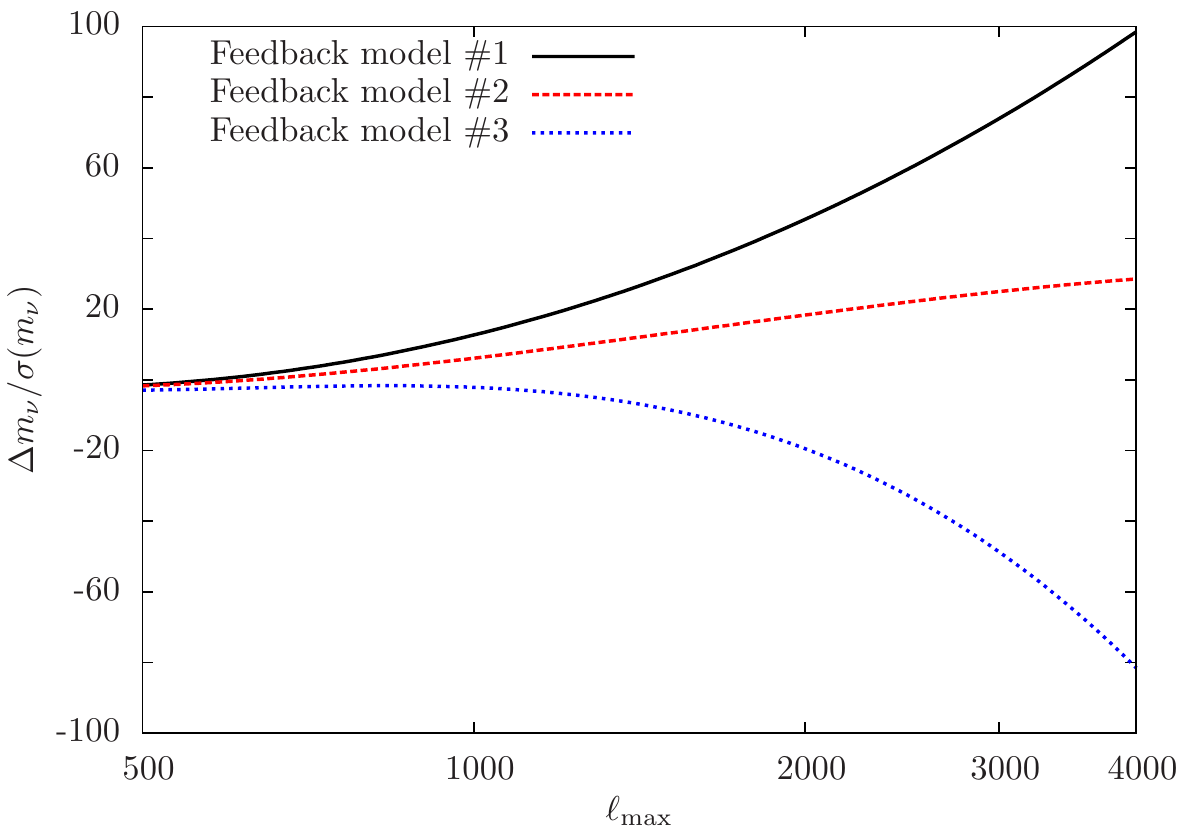}}
\end{center}
\caption{ Bias, or systematic error in the estimated neutrino mass, compared to the statistical error, for $m_\nu$ = 0.3 eV. The bias is huge compared to the statistical error, even exceeding $100\sigma(m_\nu)$  (although the results are not valid when $|\Delta m_\nu| > m_\nu$), implying that a precise measurement of $m_\nu$ is not possible unless the matter power spectrum is known to similar accuracy.
\label{fig5} }
\end{figure}

Fig. \ref{fig5} shows the bias in the neutrino mass relative to the statistical error in $m_\nu$. We see that the bias is very large (note that the formalism is invalid when $|\Delta m_\nu| > m_\nu$). This systematic error can be sufficiently large as to render cosmological weak lensing tomography an ineffective probe of neutrino mass. In each of the cases shown in Fig.~\ref{fig5}, the true neutrino mass would be ruled out at very high confidence by an analysis that neglects baryonic effects on the power spectrum. We therefore caution that experiments that aim to minimize $\sigma(m_\nu)$ must ensure that the power spectrum can also be measured to sufficient accuracy so that $\Delta m_\nu \lesssim \sigma(m_\nu)$. If we were to restrict ourselves to values of $\ell_{\rm max}$ such that the systematic error $\Delta m_\nu$ is less than the statistical error $\sigma(m_\nu)$, we find $\ell_{\rm max} < 650$ for Feedback model \#2, and even smaller for the other two models. The statistical error on these scales $\lesssim$ 0.04 eV, which is perhaps, barely sufficient to detect neutrinos in the minimal normal hierarchy. 

\begin{figure*}[!t]
\begin{center}
\scalebox{0.55}{\includegraphics{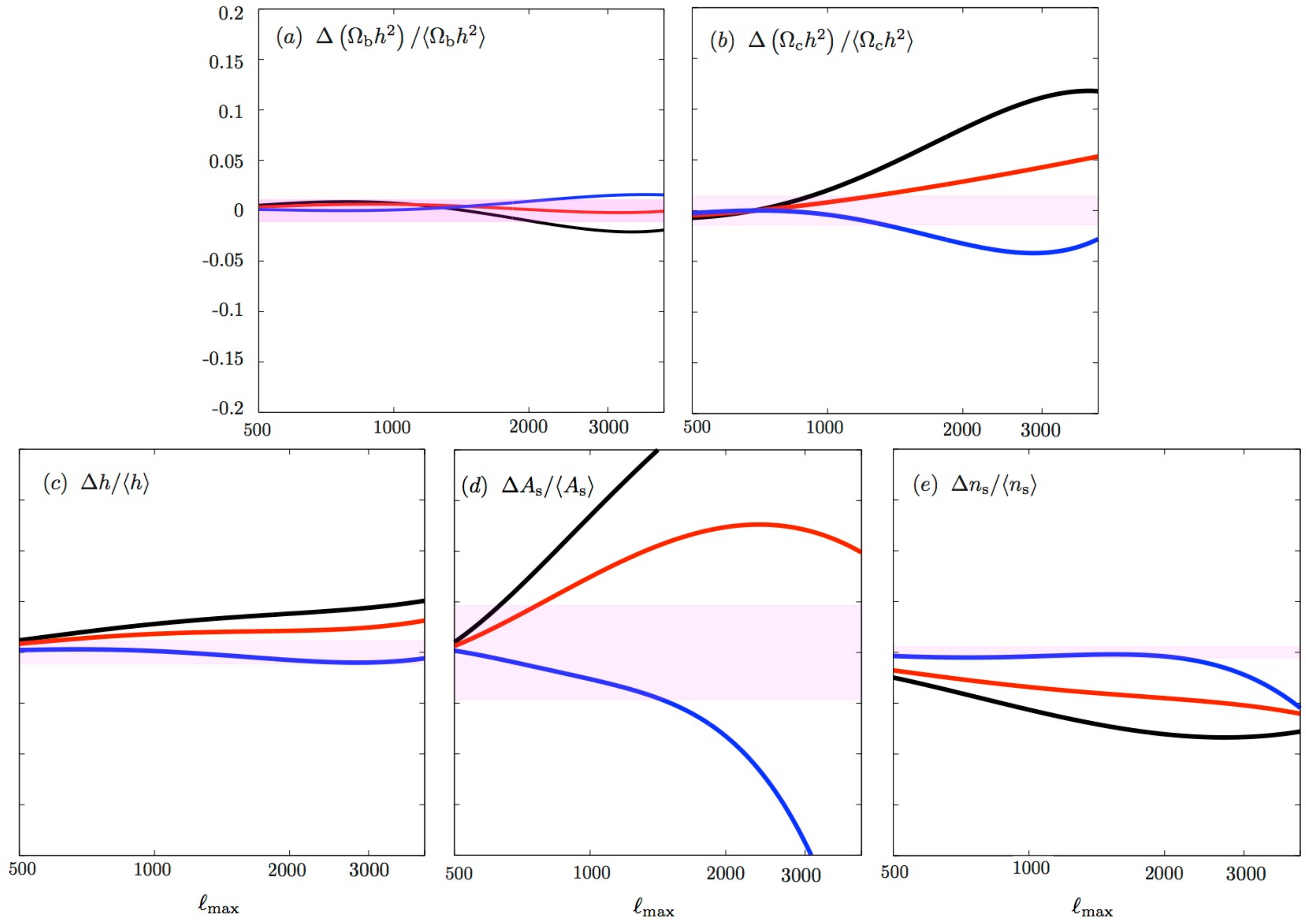}}
\end{center}
\caption{Bias in the cosmological parameters $\vec \theta = \left \{\Omega_{\rm b}h^2, \Omega_{\rm c}h^2, h, A_{\rm s}, n_{\rm s}\right\}$ (relative to the mean values) due to ignoring feedback. The shaded band shows the fractional statistical error. While $\Delta\theta_i \ll \theta_i$, the bias  is still large compared to the statistical error.
\label{fig6} }
\end{figure*}

\emph{Biases in the cosmological parameters:} Ignoring baryonic processes also introduces a bias in the other cosmological parameters, as we see from Eq. [\ref{bias}]. However, these biases are expected to be small since they are well constrained by observations of the CMB (note that current constraints on the neutrino mass from CMB+BAO are not very strong). Fig. \ref{fig6} shows the biases in all the cosmological parameters $\theta_i$ that we consider, namely $\Omega_{\rm b}h^2, \Omega_{\rm c}h^2, h, A_{\rm s}, n_{\rm s}$, for the three feedback processes. The biases $\Delta\theta_i$ are shown as a fraction of the mean value $\bar\theta_i$ obtained from CMB+BAO observations. When the observed data (which includes effects of small-scale physics) is fit to a theory model that ignores these feedback processes, the different parameters change in a non-trivial manner, in order to compensate for the effect of feedback. The scalar spectral index $n_{\rm s}$ may mimic the effect of feedback. For example, when one considers feedback models \#1 and \#2, we find a substantial damping in the power spectrum.

\begin{figure}[!t]
\begin{center}
\scalebox{0.8}{\includegraphics{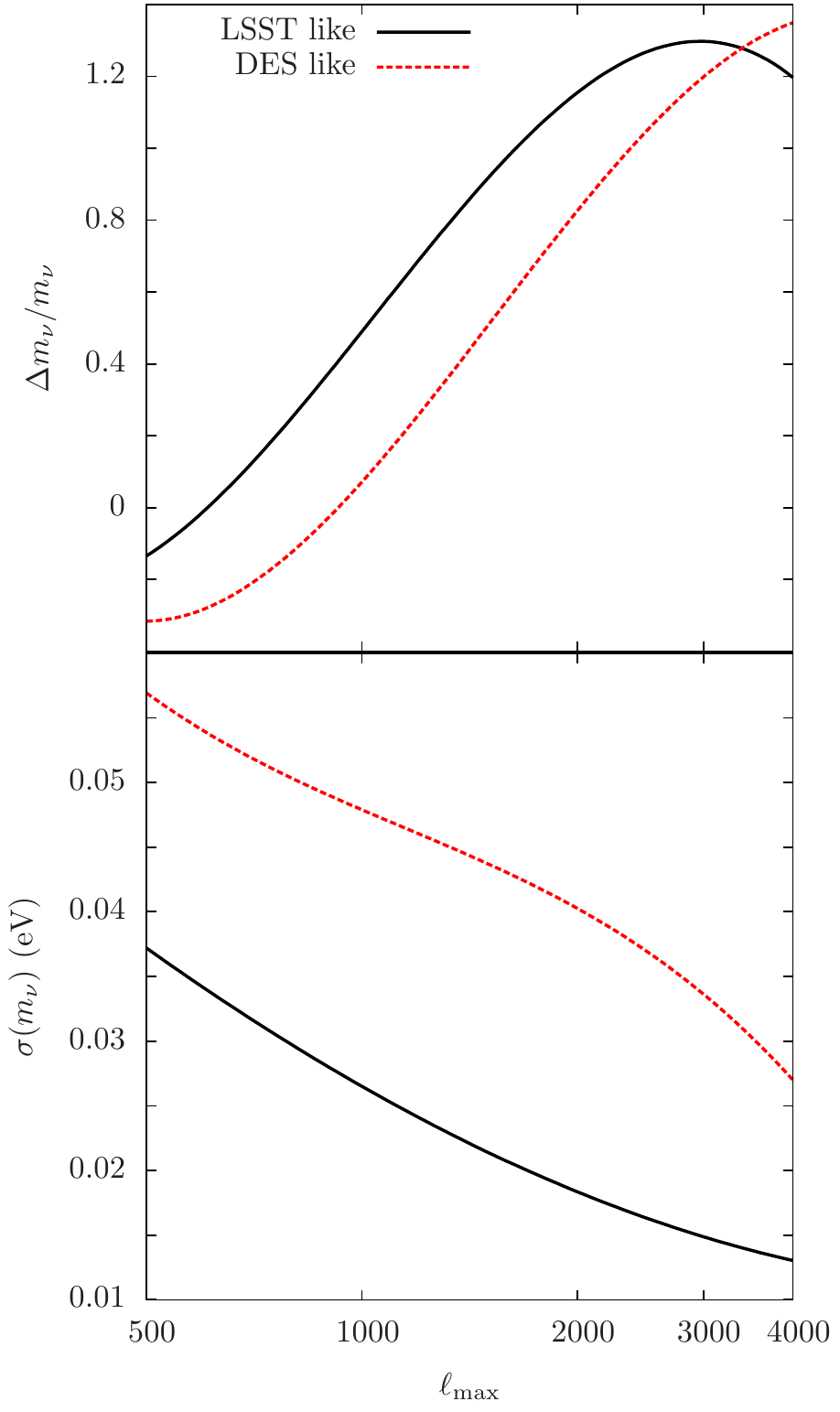}}
\end{center}
\caption{Comparison of an LSST-like survey ($z_0 = 0.92, f_{\rm sky} = 0.5$, 50 galaxies/arcmin$^2$)   with a DES-like survey ($z_0 = 0.60, f_{\rm sky} = 0.1$, 10 galaxies/arcmin$^2$), for feedback model \#2, and $m_\nu$ = 0.3 eV. 
\label{fig7} }
\end{figure}

Decreasing $n_{\rm s}$ from its mean value will result in smaller power on small scales, similar to what feedback models \#1 and \#2 predict. A similar effect occurs when the neutrino mass is allowed to vary, as we have seen. Increasing the neutrino mass causes a scale dependent damping, although neutrino masses affect large scales also. One would then need to \emph{increase} power on large scales since baryonic feedback does not damp the largest scales. The parameters $A_{\rm s}$ and $\Omega_{\rm c}h^2$ can provide such a counterbalance. We note that $A_{\rm s}$ being scale independent, is often hard to constrain. It has been shown \cite{arvi1,arvi2} that other physical  process such as  dark matter annihilation can bias $A_{\rm s}$  to higher values.


Let us now consider the effect of the survey parameters $z_0, n, f_{\rm sky}$, and $\sigma_\epsilon$ on the bias $\Delta m_\nu$ and  the statistical error $\sigma(m_\nu)$. The Dark Energy Survey (DES) which is currently collecting data, is designed to probe the origin of the acceleration of the Universe.  DES will use weak gravitational lensing as one of the probes of dark energy, and may therefore be susceptible to errors caused by uncertainties in baryonic physics. Fig. \ref{fig7} shows the performance of a DES-like survey compared to an LSST-like survey assuming $m_\nu$ = 0.3 eV, for feedback model \#2 (including galaxies in the range $0<z<2.5$ in 5 bins). For the DES-like survey, we choose the survey parameters to be (see for example, \cite{kenji}) $z_0 = 0.6, f_{\rm sky} = 0.1$ and $n$ = 10 galaxies/arcmin$^2$ (compared to $z_0 = 1.0, f_{\rm sky} = 0.5$ and $n$ = 50 galaxies/arcmin$^2$ for the LSST-like survey). In both cases, we set $\sigma_\epsilon$ = 0.22. We see that a substantial bias in the neutrino mass is still observed with the DES-like survey. We therefore emphasize that baryonic feedback effects should be included in the analysis of weak lensing data from DES.
\begin{figure}[!t]
\begin{center}
\scalebox{0.5}{\includegraphics{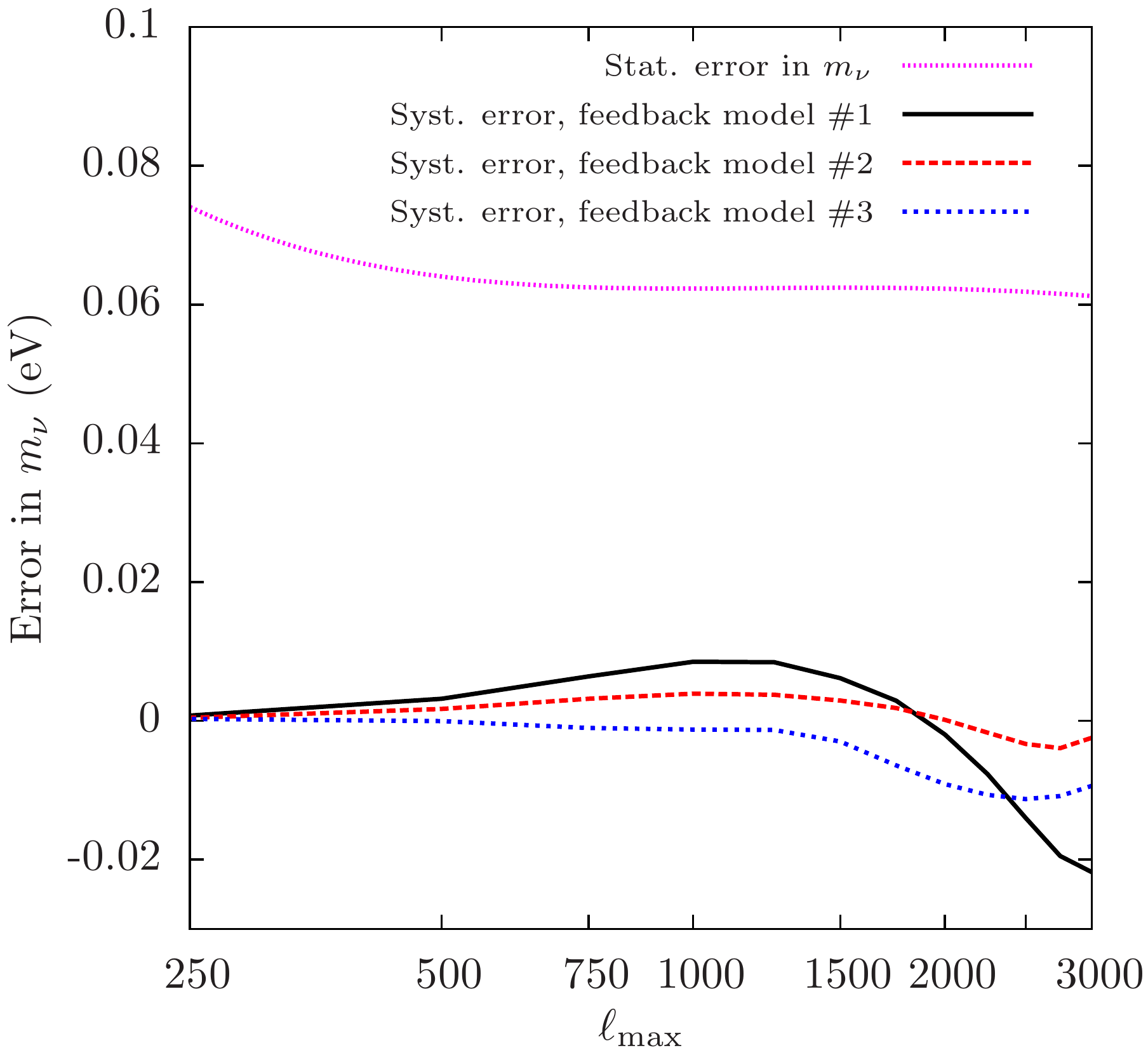}}
\end{center}
\caption{Expected errors in the neutrino mass from CMB lensing, assuming $f_{\rm sky} = 0.5$ and cosmic variance error bars. The magenta curve shows the statistical error in $m_\nu$, while the black (solid), red (dashed), and blue (dotted) curves show the systematic error due to feedback models \#1, \#2, and \#3. The true neutrino mass was set to 0.1 eV. For all models,  the systematic error is  smaller than the statistical error.
\label{fig8} }
\end{figure}

\emph{Bounds from the CMB:}  It is interesting to ask whether other cosmological probes can constrain neutrino masses as well. There exist several probes of neutrino masses such as the Lyman-$\alpha$ forest, galaxy clustering, cluster abundances, etc. A particularly promising probe is the cosmic microwave background which is being measured by current ground based experiments to very high precision on small scales. As CMB photons travel towards us they are distorted, creating secondary anisotropies. Gravitational lensing from large-scale structure is one source for these secondary anisotropies. The high precision CMB temperature measurements have lead to several experiments detecting lensing of the CMB \cite{Smith07,Das11,VanEng12,PlanckLens13}. CMB Polarization experiments also help probe lensing of the CMB. The South Pole Telescope with Polarization (SPTPol) has obtained the first detection of the polarization $B$-mode on small scales through gravitational lensing of the $E$-mode \cite{sptpol}, which may be used to reconstruct the lensing potential. The {\sc POLARBEAR} collaboration has also measured lensing in CMB polarization maps \cite{PolarB}. Future CMB experiments will be able to obtain robust constraints on neutrino masses from lensing. Authors \cite{snowmass_people} estimate that a future Stage-IV CMB experiment together with current large scale structure surveys will be able to measure neutrino masses to an accuracy of $\sigma(m_\nu) < 0.045$ eV, a bound which may be reduced further by including future galaxy surveys.

Fig. \ref{fig8} shows the errors that may be expected from a cosmic variance limited CMB lensing experiment assuming a sky fraction $f_{\rm sky} = 0.5$, and a neutrino mass $m_\nu$ = 0.1 eV. Here we use only the power spectrum of the CMB lensing potential, which is the quantity most affected by small-scale variations in the matter power spectrum. The magenta curve shows a statistical error $\approx$ 0.06 eV if multipoles up to $\ell_{\rm max} = 2000$ can be measured. The black (solid), red (dashed), and blue (dotted) curves show the systematic errors in the measurement of neutrino masses due to ignoring baryonic feedback, considering feedback models \#1, \#2, and \#3 respectively. Even in the case of AGN feedback (model \#1), the effects are small. This is because CMB lensing is more sensitive to structure at higher redshifts ($z \gtrsim 2$) and on larger scales compared to galactic weak lensing. At higher redshifts the baryon bias is small, and there are also fewer large halos. We also note that the OWLS project used $\sigma_8$ = 0.74, significantly smaller than the value measured by the Planck satellite $\sigma_8 \approx$ 0.83 \cite{planck}, which would underestimate the number of halos at higher redshifts. In all cases, the effect of baryonic processes on CMB lensing is smaller than the statistical error. Nevertheless experiments that hope to measure the sum of neutrino masses to accuracies $\sigma(m_\nu) < 0.02$ eV will need to include baryonic effects in modeling the lensing potential of the CMB. The lack of sensitivity of CMB lensing to small-scale baryonic physics makes it a  valuable probe to be used in conjunction with galactic  lensing surveys.

\section{Conclusions}

In this article, we discussed how future weak lensing surveys, such as the LSST and Euclid, can measure neutrino masses through weak gravitational lensing of large scale structure. We showed that weak lensing is sensitive to the small-scale non-linear matter power spectrum and is a powerful tool with which to probe neutrino masses. We then investigated the effect of including various baryonic feedback processes on the shear power spectrum. It was found that there is indeed a large effect.

We studied 3 feedback models using results from the publicly available Overwhelmingly Large Simulations (OWLS) project. Feedback model \#1 included AGN feedback from gas accretion on to black holes, with 15\% of the radiated energy coupling to the gas. Feedback model \#2 considered a top heavy stellar IMF which yields more supernova energy compared to the Chabrier IMF. Feedback model \#3 ignored supernova feedback as well as cooling by heavy elements.

In the case of feedback model \#3, the dominant effect is a boost in the power spectrum on small scales, due to contraction of halos in response to baryonic condensation.  With feedback models \#1 and \#2, we observed a damping of the power spectrum on small and intermediate scales due to thermal gas pressure, while the power spectrum is boosted on very small scales. Feedback model \#1 in particular shows a substantial damping of the power spectrum (nearly a 30\% effect on scales $k \sim$ 10 $h$/Mpc). It is of course important to question whether such large AGN feedback results in realistic models of large scale structure. Authors \cite{agn} studied this issue in detail and found that only simulations that include AGN feedback yielded stellar mass fractions, star formation rates, and stellar age distributions in good agreement with current observational estimates. We note the exciting possibility that future lensing surveys may be able to provide details on AGN feedback, but leave a detailed analysis to future work.

We then discussed the formalism of weak lensing and obtained error bars characteristic of the LSST experiment. In the absence of systematic errors due to small-scale physics, photometric errors, beam errors, etc, weak lensing in combination with current CMB+BAO data is a powerful probe of neutrino masses, achieving sensitivities $\sigma(m_\nu) \approx 0.01$ eV, when $N_{\rm eff}$ is held constant at its standard model value of $N_{\mathrm{eft}}=3.046$. Such precision observations can measure neutrino masses at high significance even in the case of the minimal, normal hierarchy and can determine whether or not neutrinos follow a ``normal" or ``inverted" mass hierarchy. However, this incredible sensitivity derives from exquisite measurements of lensing shear, so weak lensing results are easily affected by relatively small systematic errors.

The effects of baryons, neglected in nearly all forecasts of the power of lensing to constrain neutrino mass, introduce a potentially important systematic error. To estimate the size of this error, we studied the systematic errors induced on neutrino mass incurred by analyzing power spectra derived from the three aforementioned simulations without an explicit model for the baryonic effects. In all three of the feedback models that we studied, the bias introduced in the neutrino mass measurement exceeds both the statistical errors on the inferred neutrino mass as well as the neutrino mass itself, particularly in the case of AGN feedback. Feedback models \#1 and \#2 result in a positive bias, i.e. an overestimate of the neutrino mass, while feedback model \#3 results in a negative bias, i.e. it underestimates the neutrino mass. Thus it is crucial to account for baryonic processes in the analyses of data from future weak lensing surveys such as DES, LSST, and Euclid.

To complement galaxy lensing, we also considered the possibility of determining neutrino masses through gravitational lensing of the CMB. We showed that future cosmic variance limited surveys with $f_{\rm sky} > 0.5$ can measure neutrino masses to an accuracy $\sigma(m_\nu) \gtrsim 0.06$ eV. While this is less sensitive than galaxy weak lensing experiments, CMB lensing is nearly unaffected by feedback processes. This is because lensing of the CMB is sensitive to structure at higher redshifts, as well as larger scales compared to galaxy lensing. Agreement of mass measurements made using these two techniques will be confirmation that baryonic feedback has been correctly accounted for, and the measured masses are unbiased. More theoretical modeling and better observations are also required to understand the clustering of matter on small scales.

\acknowledgements{
We thank Rachel Mandelbaum for helpful discussions. A.N. and H.T. acknowledge funding from NASA ATP grant NNX14AB57G. H.T. acknowledges funding from Department of Energy grant DE-SC0011114, and National Science Foundation grant AST-1312991. A.N. and A.R.Z. thank the Pittsburgh Particle physics, Astrophysics, and Cosmology Center (Pitt-PACC), and the Department of Physics and Astronomy at the University of Pittsburgh, for partial financial support. A.R.Z. acknowledges funding from the National Science Foundation through grants AST 0806367 and PHY 0968888. N.B. thanks the Bruce and Astrid McWilliams Center for Cosmology for financial support. Computations were done using the {\scriptsize COMA} cluster at Carnegie Mellon University. We acknowledge the use of the Legacy Archive for Microwave Background Data Analysis (LAMBDA), part of the High Energy Astrophysics Science Archive Center (HEASARC). HEASARC/LAMBDA is a service of the Astrophysics Science Division at the NASA Goddard Space Flight Center.
}

\bibliographystyle{revtex}
\bibliography{references}

\end{document}